\documentstyle[12pt,epsf,epsfig]{article} 
\newcommand{\be}{\begin{equation}}
\newcommand{\ee}{\end{equation}}
\newcommand{\bea}{\begin{eqnarray}}
\newcommand{\eea}{\end{eqnarray}}

\newcommand{\vk}{{\bf k}}

\newcommand{\vx}{{\bf x}}
\newcommand{\vq}{{\bf q}}

\newcommand{\vep}{{\mbox{\boldmath$\epsilon$}}}

\begin{document}

\rightline{TTP97-35}
\rightline{NTZ-24/97}
\rightline{hep-ph/9709352}
\rightline{September 1997}
\bigskip
\begin{center}
{\Large {\bf Lepton pair production by a high energy  
photon in a strong electromagnetic field}}
\end{center}
%\vspace{0.2cm}
\smallskip
\begin{center}
{\large{\sc Dmitri Ivanov }}
%$^{\dagger,\$}$}}
\\
\vspace{0.4cm}
%{\sl Institut f\"{u}r Theoretische Physik}\\
%{\sl Universit\"{a}t Leipzig}\\
%{\sl D--04109 Leipzig, Germany}\\[2mm]
%$^\dagger$ 
{\sl Naturwissenschaftlich-Theoretisches Zentrum 
und Institut f\"ur Theoretische Physik,}
{\sl \\ Universit\"at Leipzig, Augustusplatz 10,}\\ 
{\sl D-04109 Leipzig, Germany}\\
{\sl and}\\
%$^{\$}$
{\sl Institut for Mathematics, 630090 Novosibirsk, Russia}
\\[2mm]
{e-mail: \sl Dmitri.Ivanov@itp.uni-leipzig.de}\\
\vspace{1cm}
%and\\
%\vspace{0.2cm}
{\large{\sc Kirill~Melnikov}}
\\
\vspace{0.4cm}
{\sl Institut f\"{u}r Theoretische Teilchenphysik}\\
{\sl Universit\"{a}t Karlsruhe}\\
{\sl D--76128 Karlsruhe, Germany}\\[2mm]
{e-mail: \sl melnikov@ttpux2.physik.uni-karlsruhe.de}\\
\vspace{1.0cm}
{\large{\bf Abstract}}\\
\end{center}
\vspace{0.1cm}
Using impact--factor representation, we consider the 
lepton pair production by an incident 
high--energy
photon in a strong electromagnetic  
field of a nucleus. By summing leading terms of 
perturbation series, we obtain  a 
simple formula for the amplitude, valid
to all orders in ${\cal O}(\alpha Z)$ 
and arbitrary field of the nucleus.
%An efficient way to obtain the total cross section of
%$\gamma A \to l^+l^- A$ is demonstrated.
Using these results,  
we derive, in a simple manner, the results for the lepton 
pair production by a virtual incident photon
in a Coulomb field.
For real incident photon our results coincide
with the known ones. 
Also, a particular example of a
non--Coulomb potential is discussed in some detail.
  
\newpage

\section {Introduction}

QED processes in the strong Coulomb field
attracted considerable attention recently. 
The understanding of such processes is important
for accurate background estimates in heavy ion collisions.
Moreover, their   analysis
is also interesting
from pure theoretical point of view. 
Indeed, the accuracy of perturbative calculations in this
case is determined by a parameter $\nu \equiv \alpha Z$,
where $\alpha $ is the fine structure constant and $Z$ is the 
charge of the heavy nucleus.
In realistic cases, $\nu \approx 1$ can be achieved and then
an exact calculations with respect to this parameter are necessary. 

The process of lepton pair production 
by a high energy ($\omega _\gamma \gg m$) photon 
in a strong Coulomb field has been considered 
in the literature 
long ago \cite {BetheMaximon} and is even 
discussed in the  text--books \cite {LL}.  
In the standard approach to this problem 
an exact in $\nu$ result is obtained by using
the Coulomb wave functions for the produced charged leptons.
Though the final result (especially for the total cross section)
is rather simple, intermediate calculations on this way are known
to be rather involved.

On the other hand, a powerful way to obtain
high energy asymptotic of  scattering amplitudes
in QED and QCD is based on the impact  representation 
for the scattering amplitude \cite {Lipatov, ChengWu.imp}. In this paper we
apply this technique  for the process of lepton pair
production by a high energy photon in a strong electromagnetic 
field of a nucleus.
Using recurrence relations
\cite{Ivanov}, which relate   
impact factors,  describing a conversion of the photon into the 
lepton pair through an exchange of
N and N+1 photons in the $t$--channel, we sum up  the leading terms
of the perturbation series in a very efficient way.
As a result we
obtain a simple formula for the amplitude valid for an arbitrary 
field of the nucleus.    

The representation of the  amplitude, 
which is derived in this paper,  is close in its meaning to a  famous 
eikonal representation for a charge--charge scattering.
It generalizes  the
eikonal representation  to the case of a dipole--charge scattering
and is valid irrespective  of an exact form of an interaction
potential between leptons and the nucleus.
This pleasant feature of our approach
permits an easy investigation of the
situation, when an interaction potential between the lepton
and the nucleus differs from the Coulomb one. This situation
emerges, for example, if one considers the nucleus as an extended
object and, consequently, describes its interaction 
with the photon by introducing a form factor 
$F(k) = \Lambda^2/(k^2+\Lambda^2)$. Below we consider
two limiting cases $\Lambda \ll m$ and $\Lambda \gg m$ in some detail.

For the Coulomb field we obtain the   
scattering amplitude and the total cross section
in the case of a lepton pair production by a virtual photon.
As a byproduct of this analysis
the known results for the lepton pair production
by a real photon are reproduced.
We feel, that the use of the representation for the 
amplitude, derived in this paper, 
considerably simplifies the calculations in this case.
This is especially notable in the case of the total cross section 
of the reaction $\gamma A \to l^+l^- A$.

The paper is organized as follows. 
In the next section we introduce our notations and
give a general discussion of the impact representation.
In Section 3  formulas
for  QED impact factors valid to all orders in ${\cal O}(\nu)$
are derived and a convenient
expression for the amplitude of the process $\gamma ^* A \to l^+l^- A$
is obtained. In Section 4 the total cross section for the
reaction $\gamma ^* A \to l^+l^- A$ is derived in a simple manner.
Section 5 is devoted to the analysis of the case of 
a non--Coulomb  potential. 
Finally we present our conclusions.

\section{Basic notations}

A typical Feynman graph which 
contributes  to the leading asymptotic
of a lepton pair
production by a high energy
photon is shown in Fig.1. Our notations are also introduced there.

%%%%%%%%%%%%%%%%%%%%%%%%%%%%%%%%%%%%%%%%%%%%%

\hspace*{3cm}

\begin{figure}[htb]
\begin{center}
\epsfig{file=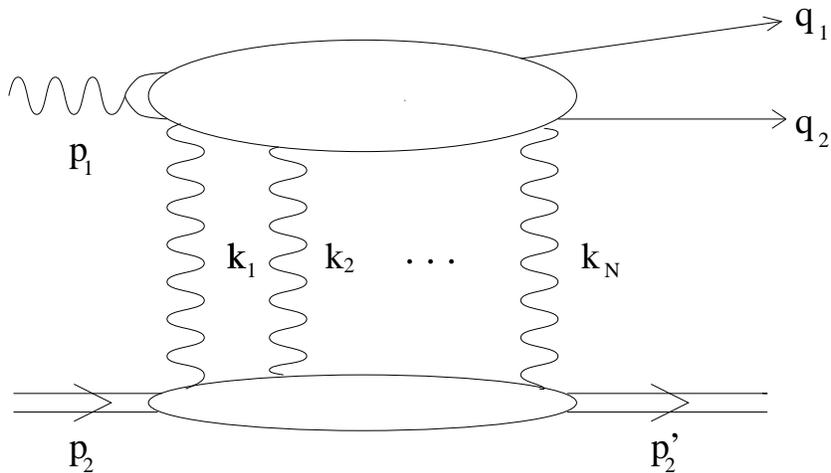,width=11cm}
\end{center}
\vspace*{0.1cm}
\caption{\label{fig1}
{\it A diagram with $N$--photons, exchanged in the $t$--channel.
Diagrams of this type contribute to the leading asymptotic
of a lepton pair production by a high energy
photon.}
%The diagram describing the process $\gamma A \to \bar l l A$.
}
\end{figure}

%%%%%%%%%%%%%%%%%%%%%%%%%%%%%%%%%%%%%%%%%%%%%%%%%%%%%%%%%%%%%%%%%%%%%%%%

The graph in Fig.1 describes an exchange of
N photons in the $t$--channel. The four momenta 
of the $t$--channel photons are denoted as $k_i$.
The upper (lower) block in Fig.1 represents an amplitude 
$A^N_{\mu_1 \dots \mu_N}$ ($B^N_{\mu_1 \dots \mu_N}$) with 
all possible attachments 
of the $t$-channel 
photons to the lepton (nuclear) line.
   
Let us denote the four momentum of the incident photon by $p_1$
and its virtuality  by $p_1^2=-Q^2$.
A transverse polarization vector
of the incident photon is  $ \vep$. 
We denote the four momenta of the produced lepton and antilepton as
$q_1$ and $q_2$, their energies as $\epsilon_1$ and $\epsilon_2$
and the corresponding Dirac spinors as $u_1$ and $u_2$. 
The mass of the lepton is $m$, its charge is $-e$.
The antilepton has the same mass and the opposite charge.
The charge of the nucleus is $eZ$, its four momentum is $p_2$.

In what follows, we consider a situation, 
when
the energy of the photon in the rest frame of the nucleus  
is  much larger than the mass of the lepton  and the virtuality of the
photon:
\be
\omega_\gamma \gg m, \sqrt {Q^2} \ .
\label{kin}
\ee
In this case the 
produced leptons are ultrarelativistic, they
 fly away in a small open angles $\theta \sim m/\omega _{\gamma}$
with respect to the momentum of the incident photon. The sum of the
energies of the produced leptons  is equal to  the 
energy of the  initial  photon
with a high  accuracy
$\sim  q^2/ M_A$.
Hence, we denote: $\epsilon_1=x\omega_\gamma$, 
$\epsilon_2=(1-x)\omega_\gamma$.

It will be convenient for us
to consider a  more general case, when the ``nucleus'' is a 
light particle with the mass equal to the mass of the 
lepton  $m$ and the charge $eZ$.
The kinematic conditions which ensure the high energy limit of the
photoproduction of the lepton pair in this case are
\be
2p_1p_2\;=\;s \gg m^2,\;Q^2.
\label{kin1}
\ee
Note, that in this limit the results for the lepton pair
photoproduction 
coincide with  the   
case of the  photoproduction on  a heavy  
nucleus at high energies\footnote{
This statement is correct for the impact factors but not for the
amplitude. The latter receives additional contributions because
of the Coulomb phases, which lepton and antilepton 
acquire in the field of the nucleus. See dedicated discussion
after Eq. (\ref {vector})}.

There are two important scales for the square of
the transverse (with respect to $p_1$ and $p_2$) 
part of the transfered momentum
$q=q_1+q_2-p_1$: $\vq^2\sim m^2,\;Q^2$ and $\vq^2\sim m^4/s,\;Q^4/s$.
We will consider mainly the first region, where the transverse momentum
is not too small:
\be
\vq^2 \sim m^2,Q^2;\; \ q^2\approx -\vq^2 \ . 
\ee 
For all diagrams of the type of 
Fig1. with the number of photons in the $t$--channel
$N\geq 2$, this region gives the leading 
contribution to the total cross section. The contribution of the small
region $\vq^2\leq q^2\sim m^4,Q^4/s$ is 
important only for the 
diagram with the exchange of one photon in the $t$--channel. 
Such diagram represents a Born process and therefore can be
easily considered separately.
 
It is convenient to introduce the
so--called Sudakov variables.
Let us define two almost light-like vectors:
\be
\tilde p_1^2=\tilde p_2^2=0 \ ,  
 \ \tilde p_1=p_1+\frac{Q^2}{s}p_2 \ ,  \ 
\tilde p_2=p_2-\frac{m^2}{s}p_1 \ .
\ee

All Minkowski vectors are decomposed  then 
in $\tilde p_1$, $\tilde p_2$ 
and  two-dimensional vectors, lying 
in the transverse plane relative to the collision axis:    
\begin{eqnarray}
k_i &=& \alpha_i  \tilde p_1 + \beta_i \tilde  p_2 + \vk_i,
\nonumber \\
q_1 &=& x_1  \tilde p_1 + y_1 \tilde  p_2 + \vq_1,
\nonumber \\
q_2 &=& x_2  \tilde p_1 + y_2 \tilde  p_2 + \vq_2.
\nonumber
\end{eqnarray}
 From Eq. (\ref{kin1}), it follows that  $x_1,\; x_2 \gg y_1,\; y_2$.
Moreover, it can be shown, that with 
the accuracy $\sim m^2/s$ 
$$
x_1=x,\;\; x_2=1-x,\;\; y_1=\frac{m^2+\vq_1^2}{xs},\;\;
y_2=\frac{m^2+\vq_2^2}{(1-x)s}.   
$$

Our aim is to sum all graphs of the type of Fig.1, 
with arbitrary number of 
photons, exchanged in the $t$--channel. 
One can show, that  
the leading contribution to the amplitude comes
from the loop momenta of the order of
$k_i^2\approx -\vk ^2_i$,
with $|\vk| \sim m,Q,|\vq|$.

In this case the  
impact representation 
(see \cite {Lipatov, ChengWu.imp})
for the  amplitude with N photons, exchanged in the $t$--channel, reads  
\be
{\cal M}_N=\frac{8\pi^2s(-i)^{N-1}}{N!} \int 
\prod ^N_{i=1} \left(\frac{{\rm d}^2 \vk _i}{(2\pi)^2\vk_i^2} \right)
\delta (\sum \limits_{i=1}^{N} {\bf k}_i - {\bf q})
J_{\gamma\to \bar l l}^N\;J_A^N. 
\label{imp.rep}
\ee

The accuracy of this equation is expected to be of the
order of $\sim m^2/s,\;Q^2/s,\;q^2/s$.
The quantities $J^N_{\gamma\to \bar l l}$ and  $J^N_{A}$
 are  called impact factors. They  describe  
the  upper and lower blocks in Fig.1. 
These impact factors  depend on the transverse 
momenta of the $t$--channel photons. Explicit formula for the
impact factors $J^N_{\gamma\to \bar l l}$ and  $J^N_{A}$ reads:
\begin{eqnarray}
J_{\gamma\to \bar l l}^N=
\int \prod ^{N-1}_{i=1} \left(\frac{{\rm d}(\beta_i s)}{2\pi i} \right)
(iA)_{\mu_1 \dots \mu_N}
\frac{ \tilde p_2^{\mu_1}\dots \tilde p_2^{\mu_N} }{s^N},
\label{impfA}
\\
J_A^N=\int \prod ^{N-1}_{i=1} \left(\frac{{\rm d}(\alpha_i s)}{2\pi i} \right)
(iB)_{\mu_1 \dots \mu_N}
\frac{\tilde p_1^{\mu_1}\dots \tilde p_1^{\mu_N}}{s^N}.
\label{impfB}
\end{eqnarray}

The impact factor of the nucleus 
$J^N_A$ is a  simple quantity, which does not depend on 
the transverse momenta of the  photons in the $t$--channel:
\be
J^N_A=i(-1)^N(eZ)^N.
\label{n.i.f}
\ee 

The expressions above are written for the  point-like ``nucleus''.
Treating the nucleus as an extended object, 
one should consider a generalization of Eqs.(5--8).
It can be easily shown, that
the only change, as compared to the point-like case, consists
in substituting the Fourier transform of the 
electric potential of the nucleus instead of each 
of the photon propagators $1/\vk _i^2$ in 
Eq. (\ref{imp.rep}).  An equivalent way to incorporate
the effects of the  non--Coulomb  field is to use Eq. (\ref{imp.rep})
with the Coulomb photon propagator $1/\vk^2$, but to consider 
 the impact factor of the nucleus $J^N_A$ as being
dependent on $t$--channel photon 
momenta. For the static field with the Fourier transform 
$F(\vk)/k^2$, this dependence is obtained by multiplying
the ``Coulomb'' impact factor of the nucleus Eq. (\ref {n.i.f}) 
by the  form factors  $F(\vk_i)$
for each of $N$ exchanged photons. 
Note that even if the nucleus can not be considered as the source of
the static field, the impact representation 
for the amplitude Eq. (\ref{imp.rep})
is still valid, but the expression for the nuclear impact factor will
be more complicated in this case. 

The central quantity for  further consideration
is the impact factor, which describes a transition of 
the incident photon to a lepton pair  
through the exchange of  N
photons in the $t$--channel.

Let us start with the simplest case --  the exchange of one photon.
For transversely polarized photon straightforward  calculations give:
\be
J^1_{\gamma\to \bar l l} 
(\vq_1,\vq_2)
= ie^2 \bar u_1 
\left[
m \hat \vep\; S^1
%(\vq_1,\vq_2) 
-2x
\left( {\bf T}^1
%(\vq_1,\vq_2) 
\vep 
\right) 
- \hat  {\bf T}^1
%(\vq_1,\vq_2)
\hat \vep 
\right] 
\frac{ \hat {\tilde p_2} }{s} u_2.
\label{1phot.i.f}
\ee
In this equation
\begin{eqnarray}
S^1 \equiv S^1(\vq_1,\vq_2) 
&= & \frac{1}{\mu^2+\vq_1^2}-\frac{1}{\mu^2+\vq_2^2}, 
\label{1phot.s}
\\
{\bf T}^1 \equiv {\bf T}^1(\vq_1,\vq_2)&=&
\frac{\vq_1}{\mu^2+\vq_1^2}+\frac{\vq_2}{\mu^2+\vq_2^2}. 
\label{1phot.t}
\end{eqnarray}
are the scalar and the vector
structures , respectively.
The quantity $\mu^2$ depends on the virtuality of the photon:
$$
\mu ^2 =m^2+Q^2x(1-x).
$$ 
We note, that for the  case of a real incident photon 
Eq.(\ref {1phot.i.f}) was first derived in  Ref.  \cite{SerKurSchil}.

In case, when the incident virtual photon is longitudinally polarized,
the one--photon impact factor reads
\be
J^1_{\gamma\to \bar l l} 
(\vq_1,\vq_2)
= -ie^2 2\sqrt{Q^2}x(1-x) 
 S^1(\vq_1,\vq_2) \bar u_1 
\frac{ \hat {\tilde p_2} }{s} u_2.
\label{1phot.i.f.l}
\ee
In deriving this equation, the polarization vector
of the photon was taken to be:
\be 
e^S_\mu =\frac{2\sqrt{Q^2}}{s}(\tilde p_2)_\mu.
\ee
Thanks to gauge invariance, the use of this expression is justified.

Impact representation for the amplitude with N photons, 
exchanged in the $t$--channel, (see Eq. (\ref{imp.rep})) 
and the results
for the 
one--photon impact factors
Eqs. (\ref{1phot.i.f},\ref{1phot.i.f.l})
are our starting points.
In the next section we use them to 
sum the leading contributions of  the graphs of the type of 
Fig1. and to derive a representation  for the scattering amplitude,
valid to all orders in ${\cal O}(\alpha Z)$.

\section{The amplitude to all orders in ${\cal O}(\nu)$}

For further calculations we need an expression for the 
impact factors
$J_{\gamma\to \bar l l}^N$. It can be shown that the $N$--photon 
impact factors are similar in their form to the one--photon 
impact factors:

\noindent a) for transversely polarized incident photon:
\be
J^N_{\gamma\to \bar l l}
(\vq_1,\vq_2)
= i(-e)^{N+1}
\bar u_1
\left [
m \hat \vep\;S^N
-2x
\left ( {\bf T}^N \vep
\right )
- \hat {\bf T}^N
\hat \vep
\right ]
\frac {\hat {\tilde p_2}}{s} u_2;
\label{Nphot.i.f}
\ee

\noindent b) for longitudinally polarized incident photon:
\be
J^N_{\gamma\to \bar l l}
(\vq_1,\vq_2)
= -i(-e)^{N+1}2\sqrt{Q^2}x(1-x)
\bar u_1
S^N
\frac {\hat {\tilde p_2}}{s} u_2.
\label{Nphot.i.f.l}
\ee

The corresponding scalar $S^N$ and vector ${\bf T}^N$ structures depend
now on the transverse momenta 
of the produced leptons 
and on the transverse momenta
of the $t$--channel photons $\{ \vk_i \},\; i=1\dots N$.
The momentum conservation constraint in this case reads:
$$
\vq_1 + \vq_2 + 
\sum \limits _{i=1}^{N} \vk_i=\vq \ . 
$$ 

It turns out (see Ref. \cite {Ivanov} for more details) 
that the  recurrence relations exist, which
relate the scalar and vector  structures, calculated  
for the exchange of $N$ photons in the $t$--channel, 
and the similar structures, calculated for the exchange 
of $N-1$ photons. These recurrence relations read:
\begin{eqnarray}
S^N(\vq _1,\vq _2,\vk _N) = S^{N-1}(\vq _1,\vq _2-\vk _N)-
S^{N-1}(\vq _1-\vk _N,\vq _2),
\\
{\bf T}^N(\vq _1,\vq _2,\vk _N) = {\bf T}^{N-1}(\vq _1,\vq _2-\vk _N)
-{\bf T}^{N-1}(\vq _1-\vk _N,\vq _2).
\end{eqnarray}
For the sake of convenience we do not indicate  
the dependence on $\vk_1\dots \vk_{N-1}$ in the above formulas.

The above recurrence relations imply, that the impact factors
$J^N_{\gamma\to \bar l l}$ vanish, if any of the $t$--channel
photons have zero transverse momentum. 
This property guarantees the infra-red finiteness of the 
impact representation for the amplitude.
Physically, this is a manifestation of the dipole
interaction of soft photons with 
the pair of leptons. For soft photon with
\be
1/|\vk|\sim 1/k >> 1/\mu \ ,
\ee  
the strength of the interaction with the lepton pair
is $\sim ek/\mu$. 
Therefore, when all $N$ photons in the $t$--channel
are soft,
the scalar and vector structures appear to 
be strongly suppressed:
%The recurrence
%relations imply, that
%these impact factors vanish, if  
%any of the $t$--c $\vk_i$  tends to zero. This property is related 
%with guage invariance and guarantees the infrared finitness of the impact
%representation (\ref{imp.rep}) for the amplitude. One more important 
%property  of these impact factors is the dipol behaviour  in the region
%of small photon momenta. 
%When the wave lenghts of all the 
%$t$--channel photons are 
%mush more then the typical transverse separation between the leptons
%\be
%1/|\vk_i|\sim 1/k >> 1/\mu \ ,
%\ee  
%it can be seen that scalar and vector structures are small 
\be
S^N \sim  {\cal O}\left(\frac{1}{\mu^2}\left( \frac{k}{\mu} \right )^N\right)
\ , \
 {\bf T}^N \sim {\cal O}\left(\frac{1}{\mu}
\left( \frac{k}{\mu} \right)^N\right) \ .
\ee
%The origin of this smallness is very clear. The photon with large wave 
%lenght interacts not with the charge of a separate lepton but with the 
%dipol momentum of a pair. Therefore each of such photon brings the small
%factor $\sim \frac{k}{\mu}$ in the amplitude.  
 
We note also,  that the impact factors $J^N_{\gamma\to \bar l l}$ are 
symmetric under the permutations of the momenta of the 
$t$--channel photons.
This property follows from the Bose symmetry and the 
fact, that all $t$--channel photons  have identical polarization
(cf. Eq. (\ref{impfA})).

The derivation of the recurrence relations  
will be presented elsewhere \cite{Ivanov}. 
We note, that for the cases $N = 2, 3$
the impact factors were earlier calculated by direct 
integration of the Feynman diagrams using 
 Eq. (\ref{impfA}) ( see \cite {GPS,GIv}).
Using recurrence relations, 
these results can be easily reproduced.

The knowledge of the recurrence relations is the crucial point 
for the subsequent analysis. 
Using them, we sum up the graphs of the type of Fig.1.
Below we  discuss the resummation for the 
scalar structure only, the derivation
for the vector structure follows along the same lines.

To calculate the amplitude for the lepton pair production we need
to calculate the integrals over 
$t$--channel loop momenta.
For this purpose we introduce:
\be
J^N_S(\vq_1,\vq_2) =\int \prod \limits _{i=1}^{N} \frac {{\rm d}^2 \vk _i}{k_i^2}  
\delta (\sum \limits_{i=1}^{N} {\bf k}_i - {\bf q})  S^N.
\label{ttt}
\ee

Using recurrence relations we find: 
\be
J^N_S(\vq_1,\vq_2) =\int  \frac {{\rm d}^2 \vk }{\vk ^2}        
(J^{N-1}_S(\vq_1,\vq_2-\vk)-J^{N-1}_S(\vq_1-\vk,\vq_2)).
\ee

It is advantageous here to write this relation for the Fourier
transform of the function $J^N_S(q_1,q_2)$ which is defined
through:
\be
J^N_S(x_1,x_2)=\frac {1}{(2\pi)^2}
\int e^{i\vq _1\vx _1 + i\vq _2\vx _2}J^N_S(q_1,q_2)
{{\rm d}^2 \vq _1}{{\rm d}^2 \vq _2} \ .
\ee
In this representation, the recurrence relations are
greatly simplified:
\be
J^N_S(x_1,x_2)=J^{N-1}_S(x_1,x_2)\; 2\pi\;\log \left (
\frac{x_1}{x_2} \right ).
\label{FTrecrel}
\ee

The boundary values $J^1_S(x_1,x_2)$ can be derived and read
\be
J_{S}^1(x_1,x_2)=K_0(\mu |x_1-x_2|)\cdot \log \left
(\frac{x_1}{x_2}\right ),
\ee
where the function $K_0(x)$ is the modified Bessel function.

Using then Eq. (\ref {FTrecrel}), one easily finds that:
\be
J_S^{N}(x_1,x_2) = \frac {K_0(\mu |x_1-x_2|)}{2\pi}\;
\left[ 
\pi
\log  \left( \frac{x_1^2}{x_2^2} \right ) 
\right]^N.
\label {jsN}
\ee

Final results for the scalar and the vector structures
are calculated then by taking the sum over $N$. We define therefore:
\be
J_S(x_1,x_2) = \sum \limits _{i=1}^{\infty} 
\frac {(-i\nu/\pi)^{N-1}}{N!}J^N_S(x_1,x_2).
\ee

Using Eq. (\ref {jsN}) to perform the summation and taking inverse
Fourier transform, we arrive at the final result for the scalar and 
the vector structures:
\be
J_S(q_1,q_2) = \frac {i}{(2\pi)^2 2\nu} \int 
{\rm d}^2 \vx _1 {\rm d}^2 \vx _2\; e^{-iq_1x_1-iq_2x_2} 
K_0(\mu |\vx _1- \vx _2|)
\left [ \left (\frac {x_1^2}{x_2^2}\right )^{-i\nu} - 1 \right ],
\label {scalar}
\ee
\be
{\bf J}_T(q_1,q_2) = \frac {-1}{(2\pi)^2 2 \nu} \int 
{\rm d}^2 \vx _1 {\rm d}^2 \vx _2\; e^{-iq_1x_1-iq_2x_2} 
\frac {\mu \left ({\bf x}_1 - {\bf x}_2 \right )}{|\vx _1 - \vx _2|} 
K_1(\mu |\vx _1-\vx _2|)
\left [ \left (\frac {x_1^2}{x_2^2}\right )^{-i\nu} - 1 \right ].
\label {vector}
\ee

For the sake of completeness, we finally present
the amplitude of the process
$\gamma ^* A \to l^+ l^- A$, expressed through
the scalar and vector impact factors. We note, however,
that the impact representation (Eqs. (5-7)) leaves unaccounted
the effect of large longitudinal distances which 
are responsible for a Coulomb phase of lepton and 
antilepton in the field of the nucleus. This classical
effect has been the subject of a rigorous treatment
in the past, its comprehensive discussion in QED can be
found in Ref. \cite {Yennie}.

According to \cite {Yennie}, the effect of soft photons
is described by  an universal 
exponential factor.
Usually, this factor is known to be 
infra--red divergent. In our case, however, the infra--red
safe {\em difference} of the Coulomb phases of lepton and antilepton
emerges. We note, that this finite answer is sensitive 
to the way one treats the ``heavy nucleus''. In fact, one
gets two different results, considering the nucleus
as infinitely heavy particle (the situation of actual
interest for us) or as a particle with the mass equal to lepton
masses (the technical trick used in this paper).
Therefore, the formulas below apply to the case of 
infinitely heavy nucleus only.
In this case, we can use the results for the exponential
factor as given in \cite {Yennie}.  
It can be checked that the leading contribution
to the soft photon exponential factor comes from the
region where $s\alpha _i \beta _i \sim \vk _i^2$.
The contribution from this 
region can therefore  be separated from the contribution 
of the region  $s\alpha _i \beta _i \ll \vk _i^2$  which 
is responsible for hard diffractive scattering
and which is correctly described by impact representation. 
For this reason,
the exponential factor and the rest of the amplitude,
described by impact factors, can be considered
separately. The result is then given by a simple multiplication.

We find the following
expression for the amplitude:\\
\noindent a) for transversely polarized incident  photon:
\be
{\cal M} = 8\pi e \;\nu \;s  \left ( \frac {x}{1-x} \right ) ^{-i\nu}
\bar u \left [m \hat \vep\;J_S(q_1,q_2) 
-2x\left ( {\bf J}_T(q_1,q_2) \vep \right ) - \hat {\bf J}_T(q_1,q_2)
\hat \vep \right ] \frac {\hat p_2}{s} u.
\label {amplitude}
\ee

\noindent b) for longitudinally polarized incident  photon:
\be
{\cal M} = -16\pi e \;\nu \;s  
 \left ( \frac {x}{1-x} \right ) ^{-i\nu}
\sqrt{Q^2}x(1-x)
\bar u J_S(q_1,q_2) 
\frac {\hat p_2}{s} u.
\label {amplitude.l}
\ee

The generalization to  the case of a non--Coulomb potential is straightforward.
Changing the photon  propagator $1/\vk^2\to F(\vk)/\vk^2$, 
where $F(\vk)$ is the electromagnetic form factor
of the nucleus,
one can repeat the derivation.  
One finds, that in this case 
Eqs. (\ref{amplitude},\ref{amplitude.l}) are still valid, if
the  vector and scalar structures are modified:
\be
J_S(q_1,q_2) = \frac {i}{(2\pi)^2 2\nu}  
\int 
{\rm d}^2 \vx _1 {\rm d}^2 \vx _2\; e^{-iq_1x_1-iq_2x_2}
K_0(\mu |\vx _1-\vx _2|)
\left [e^{-i\nu\varphi \left (\vx_1,\vx_2 \right )} - 1 \right ],
\label {scalarnc}
\ee
\be
{\bf J}_T(q_1,q_2) = \frac {-1}{(2\pi)^2 2\nu}  
\int 
{\rm d}^2 \vx _1 {\rm d}^2 \vx _2\; e^{-iq_1x_1-iq_2x_2}
\frac {\mu (\vx _1 -\vx _2 )}{|\vx _1 - \vx _2 |}
K_1(\mu |\vx _1-\vx _2|)
\left [e^{-i\nu\varphi \left (\vx_1,\vx_2 \right )} - 1 \right ].
\label {vectornc}
\ee
where
$$
\varphi \left (\vx_1,\vx_2 \right ) = \frac {1}{\pi} \int 
\frac {{\rm d}^2 \vk }{k^2}\;F(\vk) \left (e^{i\vk \vx _2}
-e^{i\vk \vx _1} \right ) \ .
$$
%Note, that in  case of spherical symmetric potential the phase
%$\varphi \left (\vx_1,\vx_2 \right )$ depends only on the lengths
%of the vectors $\vx_1$ and $\vx_2$. 
For the Coulomb potential $F(\vk)=1$. Then 
\be
\varphi \left (\vx_1,\vx_2 \right )=\varphi_C \left (x_1,x_2 \right )=
\log{\frac{x^2_1}{x_2^2}} \ ,
\ee
and Eqs. (\ref{scalar},\ref{vector}) are reproduced.

A few words about  helicity amplitudes are in order here. 
%Representation
%(\ref{amplitude}) for the amplitude seems to us very convinient. 
Using
the explicit form for the Dirac spinors
in Eq. (\ref {amplitude}), one can show that
the parts 
of the amplitude which are proportional to the scalar and vector structures,
correspond to the helicity amplitudes with total helicity of the pair
$\pm 1$ for $J_S$  and $0$ for ${\bf J}_T$. 
In the impact parameter 
representation,  helicity coincides
with the
projection of a spin on the collision axis. 
Therefore, in the case of a  spherically symmetric potential, 
$J_S(\vx_1,\vx_2)$ and ${\bf J}_T(\vx_1,\vx_2)$
must correspond to the eigenfunctions of the 
projection of the total 
angular momentum on the collision axis with the eigenvalues $0$ and 
$\pm 1$, respectively. 
The   correctness of this statement can be seen directly from 
Eqs. (\ref{scalarnc},\ref{vectornc}).

%Usin obtained above representation for the amplitude 
%we derive in the next section the results for the amplitude and the total 
%cross section for the case of pure  Coulomb  nuclear potential.

\section{Coulomb potential}

\subsection{The amplitude}
 
The formulas presented in the previous section provide a
basis for further calculations. 
In what follows,  
we consider the scalar structure $J_S$ only. The calculation
for the vector structure is 
easily performed along the same lines. 

We start with substituting the Fourier transform for the 
Bessel function
$K_0(\mu |x_1-x_2|)$ in to Eq. (\ref {scalar}). 
Integrating over $x_1$ and $x_2$, we arrive at the following
expression for $J_S$:
\be
J_S(q_1,q_2) = \frac {-2i\nu}{(2\pi)} \int 
\frac {{\rm d}^2 \vk }{(\vk ^2+\mu ^2)((\vk-\vq _1)^2)^{1-i\nu} 
((\vk+\vq _2))^{1+i\nu}}.
\ee
Further integrations can be easily performed using powerful
methods developed for calculating one loop Feynman integrals.
Introducing Feynman parameters and doing the projective
transformation \cite {HV}, we rewrite $J_S$ as:
\be
J^S(q_1,q_2) = \frac {i\nu}{|\Gamma (1-i\nu)|^2} \frac {1}{\mu^2q^2}
\left (\frac {\xi _2}{\xi _1} \right )^{\i \nu} \left [
J_0 + (\xi _1 -1) J_1 +(\xi _2 -1) J_2 \right ],
\ee
where
\bea
J_0 &=& \int \frac {{\rm d}u_1{\rm d}u_2{\rm d}u_3 \delta (1-\sum u_i)
 u_1^{-i\nu} u_2^{i\nu}}{\left(u_3 + \delta u_1 u_2 \right )^2},
\\ \nonumber
J_1 &=& J_2^* = \int \frac {{\rm d}u_1{\rm d}u_2{\rm d}u_3 \delta (1-\sum u_i)
 u_1^{1-i\nu} u_2^{i\nu}}{\left(u_3 + \delta u_1 u_2 \right )^2},
\eea
and
\be
\delta = \frac {q^2}{\mu ^2}\xi _1 \xi _2,~~~~~
\xi _{1,2} = \frac {\mu ^2}{q_{1,2}^2 + \mu ^2}.
\ee

These integrals are rather simple and can be calculated directly.
As an example, we quote the result for $J_0$:
\be
J_0 = \frac {|\Gamma (1-i\nu)|^4}{i\nu} \left \{
F(2-i\nu,1+i\nu;2;z) - (1+i\nu) \delta 
F(2-i\nu, 2+i\nu;2;z) \right \}.
\ee
Here $z = 1-\delta$ and $F(a,b;c;z) \equiv \mbox{}_2F_1(a,b;c;z)$
is the Gauss hypergeometric function.

The hypergeometric functions can be simplified by using Gauss
relations and the formula
\bea
F(a,b+1;c+1;z) &=& \frac {c}{c-a}F(a,b;c;z) +
\frac {c(1-z)}{b(a-c)}F'(a,b;c;z),
\\ \nonumber
F'(a,b;c;z) &=& \frac {{\rm d}}{{\rm d}z}F(a,b;c;z).
\eea

Finally, we get the following results  for the scalar and the 
vector structures:
\bea
J_S(q_1,q_2) &=& \frac {|\Gamma (1-i\nu)|^2}{\mu ^2\;q^2}
\left (\frac {\xi _1}{\xi _2} \right )^{-i\nu} \left \{
(\xi _1 - \xi _2) F(z) - \frac {i\delta}{\nu}(\xi _1 + \xi _2 -1) F'(z)
\right \},
\label {scexact}
\\ 
{\bf J}_T(q_1,q_2) &=& \frac {|\Gamma (1-i\nu)|^2}{\mu ^2\;q^2}
\left (\frac {\xi _1}{\xi _2} \right )^{-i\nu} \left \{
(\xi _1 {\bf q}_1 + \xi _2 {\bf q}_2) F(z) - \frac {i\delta}{\nu}
(\xi _1 {\bf q}_1 - \xi _2 {\bf q}_2) F'(z)
\right \},
\label {veexact}
\eea
where the function $F(z)$ reads:
\be
F(z) =F(i\nu,-i\nu;1;z).
\ee

Inserting Eqs. (\ref{scexact},\ref{veexact}) into Eqs.  
(\ref{amplitude},\ref{amplitude.l}) one  obtains a final expression 
for the amplitudes of the process $\gamma^* A\to  l^+ l^- A$
in the strong Coulomb field.
For the lepton pair production by a real  ($Q^2=0$) 
incident photon, 
the expression for the amplitude 
coincides with the known result \cite{LL}.

\subsection {Total cross section}

Using the representation for the amplitude  Eq. (\ref {amplitude}), 
one derives the result for the
differential cross section of the $\gamma^* A \to l^+ l^- A$:
\be
d\sigma = \frac{2mr_e\nu^2}{\pi^2}
\left [
m^2|J_S|^2+|{\bf J}_T|^2(x^2+(1-x)^2)
\right ]
{\rm d}x{\rm d}^2\vq _1{\rm d}^2\vq _2,~~~
r_e = \frac {\alpha }{m}.
\label{crossection}
\ee

At this point it is convenient to separate a part of the cross section
originating from the exchange of one photon $d\sigma _1$
and the other part  $d\sigma _2$, which describes the contribution 
of the diagrams Fig.1 with $N\geq 2$  and their interference with 
the one photon exchange diagram. 
For this purpose we write:
\be
d\sigma = d\sigma _1 + d\sigma _2.
\ee

The one--photon exchange  
cross sections can be calculated without much problems.
For transversely polarized incident photon, the result
 is given by the following equation:
\be
\sigma _1^T = \frac {4}{3}Z^2 \alpha r_e^2 \int \limits  _{0}^{1}
{\rm d}x \left \{ \frac {1}{g(x,z)^2}
+ \frac {2\left (x^2+(1-x)^2 \right )}{g(x,z)} \right \}
\left \{ \log \left (\frac {2x(1-x)\omega}{m\sqrt{g(x,z)}} \right)
- \frac {1}{2} \right \},
\ee
where
$$
g(x,z) = 1+zx(1-x),~~~~z = \frac {Q^2}{m^2}.
$$

For longitudinally polarized incident photon the cross section reads:
\be
\sigma _1^S =  \frac {4}{3}Z^2 \alpha r_e^2 \int \limits  _{0}^{1}
{\rm d}x  \frac {4zx(1-x)}{g(x,z)^2}
\left \{ \log \left (\frac {2x(1-x)\omega}{m\sqrt{g(x,z)}} \right)
-\frac {1}{2} \right \}.
\ee

The integration over $x$ can be done
in the above formulas, leading to
complicated expressions involving dilogarithms. We do not quote these
results here. 

In the case of real incident photon, the integration over
$x$ is easily performed, leading to the 
formula for $\sigma _1$, first obtained by 
Bethe and Heitler
\cite {BH}:
\be
\sigma _1 = \frac {28}{9}\; r_e^2\; Z^2 \alpha \; \left (
\log \left (\frac {2\omega}{m} \right ) - \frac {109}{42} \right ).
\label {1phcr}
\ee

A real challenge in the standard calculations is to get an expression
for $d\sigma _2$. 
A common way to calculate $\sigma _2$ is to use the
formulas for the Coulomb wave functions in the ultrarelativistic 
limit. In essence, this is equivalent to use our expressions
for the scalar and vector structures Eqs. (\ref {scexact}, \ref{veexact}).
This is not, however, the most efficient way. It turns out that
for the derivation of the total cross section the expressions
for $J_S$ and ${\bf J}_T$, as given by Eqs. (\ref {scalar},\ref {vector}),
are more convenient. 

For transversely polarized incident photon, 
using Eq. (\ref {crossection}), one obtains:
\be
\sigma_{2}^T=\frac{2 mr_e\nu^2}{\pi^2}
\int\limits_0^1 {\rm d}x \left [
m^2\;A_1 + (x^2+(1-x)^2)\;A_2
\right ].
\label {sigma2}
\ee
For longitudinally polarized incident photon the result is
\be
\sigma_{2}^S=\frac{2 mr_e\nu^2}{\pi^2}
\int\limits_0^1 dx \left (4Q^2x(1-x)   
A_1
\right ).
\label {sigma2.s}
\ee
In Eqs. (\ref {sigma2}, \ref {sigma2.s}) we used:
\bea
A_1 &=& \int {\rm d}^2 \vq _1 {\rm d}^2 \vq _2 \left ( |J_S|^2 -
|J_S^{\rm one-photon}|^2 \right ),
\\
A_2 &=& \int {\rm d}^2 \vq _1 {\rm d}^2 \vq _2 \left ( |{\bf J}_T|^2 -
|{\bf J}_T^{\rm one-photon}|^2 \right ).
\eea

Using the representation for the scalar and vector impact factors
given in Eqs. (\ref {scalar},\ref {vector}) and integrating
over $\vq _1$ and $\vq _2$, we obtain:
\bea
A_1 = \frac {1}{4 \nu^2}
\int {\rm d}^2 \vx_1{\rm d}^2 \vx_2 K_0^2(\mu |x_1-x_2|) 
\left \{ \left [\left(\frac {x_1^2}{x_2^2} \right )^{i\nu} -1 \right ]
\left [ \left(\frac {x_1^2}{x_2^2} \right )^{-i\nu} -1 \right ]
 - \nu^2 \log \left (\frac {x_1^2}{x_2^2} \right ) \right \},
\nonumber \\
A_2 = \frac {\mu^2}{ 4 \nu^2}
\int {\rm d}^2 \vx _1{\rm d}^2 \vx _2 K_1^2(\mu |x_1-x_2|) 
\left \{ \left [\left(\frac {x_1^2}{x_2^2} \right )^{i\nu} -1 \right ]
\left [ \left(\frac {x_1^2}{x_2^2} \right )^{-i\nu} -1 \right ]
 - \nu^2 \log \left (\frac {x_1^2}{x_2^2} \right ) \right \}.
\nonumber \\
\eea

It is convenient  to introduce
new variables for further integration: 
${\bf R} = ({\bf x}_1+{\bf x}_2)/2$,
${\bf r} = {\bf x}_1 - {\bf x}_2$ 
and ${\bf n} = {\bf r}/r$. Integrations over ${\bf R}$ and 
${\bf r}$ then completely factorize. One gets therefore:

\bea
A_1 &=& \frac {\pi}{2\nu^2 \mu ^4}
\left (\int {\rm d} x x^3 K_0^2(x) \right )\;I_{\nu},~~~~
A_2 = \frac {\pi}{2\nu^2 \mu ^2}
\left ( \int {\rm d} x x^3 K_1^2(x) \right )\;I_{\nu},~~~~
\nonumber \\
I_{\nu} &=&
\int {\rm d}^2 {\bf R} 
\left \{ 2 - \left(\frac {R^2}{({\bf R}-{\bf n})^2} \right
)^{i\nu}- \left(\frac {R^2}{({\bf R}-{\bf n})^2} \right
)^{-i\nu} - \nu^2 \log^2 \left( \frac {R^2}{({\bf R}-{\bf n})^2} 
\right) \right \}.
\eea

The calculation of $I_{\nu}$ can be done especially  
simple by introducing 
dimensional regularization to calculate it term by term. We note that
in this case the first (constant) term  in the integrand of $I_{\nu}$
does not contribute at all.

One gets:
\be
I_{\nu} = -2\pi \nu ^2 
 \left \{ \Psi (1-i\nu) + \Psi (1+i\nu) - 2\Psi(1)
\right \},
\ee
where $\Psi(z) = {\rm d}\Gamma (z) /{\rm d}z$.

Using also
$$
\int {\rm d}x x^3 \left (K_0^2 (x),\; K_1 ^2 (x) \right ) = 
\left (\frac {1}{3},\; \frac {2}{3} \right ),
$$
we obtain the result for the $\sigma _2^T$:
\be
\sigma _2^T \left( z \right )= \sigma _2^T (0)\;
G_T\left (z \right ),~~~~z = \frac {Q^2}{m^2}.
\ee
In this equation
\be 
G_T(z)=\frac {3}{7} \int\limits^1_0
{\rm d}x \frac{1+2(x^2+(1-x)^2)g(x,z)}{g(x,z)^2}
\ee  
and
\be
\sigma _2^T (0) = -\frac {28}{9} Z^2 \alpha r_e^2 f(\nu),~~~~
f(\nu)=\frac{1}{2}\left [ \Psi (1-i\nu) + \Psi (1+i\nu) - 2\Psi(1)
\right ].
\label{sigma2L}
\ee

%An explicit  expression for the function $G_T(z)$ reads:
%\be
%G_T(z) = \frac {6}{7} \left \{
%-\frac {\left (8+z \right )}{z(z+4)} 
%+\frac {\left (16+14z+2z^2 \right)}{ z^{3/2}(z+4)^{3/2}} 
%\log \left (\frac {\sqrt {z+4} + \sqrt {z}}{\sqrt {z+4} - \sqrt {z}}
%\right ) \right \}.
%\ee

>From the above formulas one easily obtains the 
result for the total cross section of the 
lepton pair production induced by a real photon. Indeed,
combining $\sigma _1$ and $\sigma _2^T(0)$ together, we obtain the 
standard result
for the total cross section of the reaction $\gamma A \to l^+ l^- A$
in the Coulomb field:
\be
\sigma = \sigma _1 + \sigma _2 = 
\frac {28}{9}\; Z^2 \alpha r_e^2\; \left (
\log \left (\frac {2\omega}{m} \right ) - \frac {109}{42} 
-f(\nu) \right ).
\ee

The result for the longitudinally polarized incident photon reads:
\be
\sigma _2^S (z) = -\frac {16}{3} Z^2 \alpha r_e^2 f(\nu) G_S(z)\ ,~~~~
G_S(z)=z\int\limits^1_0
{\rm d}x \frac{x(1-x)}{g(x,z)^2}
\label{sigma2S}
\ee

%For the transversly polarized virtual incident photon,
%the $\sigma _2$, depends 
%on the virtuality of the photon.
%One finds that
%\be
% \sigma _2^T \left( \frac {Q^2}{m^2} \right )= \sigma _2 (0)^T\;
%f\left ( \frac {Q^2}{m^2} \right ),
%\ee
%where
%\be 
%\ f(u)=\int\limits^1_0
%dx \frac{3}{7}\frac{1+2(x^2+(1-x)^2)(1+ux(1-x))}{(1+ux(1-x))^2}
%\ee  
%\be
%f(x) = \frac {6}{7} \left \{
%\frac {\left (16 +11x+2x^2 \right )}{x(x+4)} 
%-\frac {\left (32+18x+2x^2 \right)}{ x^{3/2}(x+4)^{3/2}} 
%\log \left (\frac {\sqrt {x+4} + \sqrt {x}}{\sqrt {x+4} - \sqrt {x}}
%\right ) \right \}
%\ee

\section{An example of  a non--Coulomb potential}

A heavy nucleus is an extended object with the inverse
radius
\be
\Lambda \sim 30 \mbox{ MeV }. 
\ee 
At such scales, the electric field of the nucleus
differs substantially from the Coulomb behavior.
Note, that the realistic values of $\Lambda $ fall into the
interval which can be described by an inequality:
$$
m_e \ll \Lambda \ll m_\mu,
$$
where $m_\mu$ and $m_e$ are muon and electron masses, respectively.
Therefore, for practical purposes, it is necessary
to consider 
two limiting 
cases: small or large $\Lambda$ in comparison with the masses of the
produced fermion. 
Below these cases are considered in turn.
In what follows, we will assume a particular form 
of the electromagnetic form factor of the nucleus:
\be
F(k)=\frac{\Lambda^2}{\Lambda^2+k^2} \ .
\label {formfactor}
\ee

In this section we also restrict our consideration to
the case of a real incident photon. We note, however, that 
the results for the virtual incident photon require
a simple rearrangement of the formulas presented below
and thus can be readily obtained if need arises. 

\subsection{ The case $m \ll \Lambda$}
 
In this case  a perturbation theory in $m/\Lambda$ 
can be developed. 
For the realistic case $m=m_e$ the non--Coulomb correction 
$\sim m^2/\Lambda^2$ is small for the pair 
production by a real photon.
We note, however, that 
this correction can become 
important for the photoproduction 
by a virtual photon, where the expansion parameter
is $\mu^2/\Lambda^2$. 

For the particular choice of the form factor
Eq. (\ref {formfactor}),
the phase $\varphi$ in Eqs. (\ref {scalarnc}, \ref {vectornc})
can be written as
\be
\varphi  = \varphi_C  - \delta \varphi,
\ee
\be
\varphi _C = 
\frac {1}{\pi} \int 
\frac {{\rm d}^2\vk}{k^2}\;\left (e^{i\vk\vx_2} -e^{i\vk\vx_1} \right ),~~~~
\delta \varphi = 
\frac {1}{\pi} \int 
\frac {{\rm d}^2\vk}{k^2+\Lambda^2} 
\;\left (e^{i\vk\vx_2} -e^{i\vk\vx_1} \right ).
\ee

As was shown above, the total cross section is
expressed through the squares of the scalar and vector impact factors,
integrated over transverse momenta of both leptons:
\bea
A^S &=& \int {\rm d}^2\vq _1 {\rm d}^2 \vq _2 |J^S(q_1,q_2)|^2,
\\
A^T &=& \int {\rm d}^2\vq _1 {\rm d}^2 \vq _2 
|{\bf J}^T(q_1,q_2)|^2.
\eea

By $\delta A^{S,T}$ we denote a correction to these quantities 
caused by 
a deviation of the interaction potential from the Coulomb one,
in the limit $\Lambda \gg m$. To calculate
these corrections we note that the ${\cal O}(m^2/\Lambda ^2)$ order
effect originates from the configurations in the coordinate space
where either $x_1 \sim \Lambda ^{-1}, x_2 \sim m^{-1}$, or
$x_2 \sim \Lambda ^{-1}, x_1 \sim m^{-1}$, in this case 
one of the particles flies 
through the nucleus.  Note that for the vector structure $A^T$
an additional contribution of the same order exists, which originates from 
the configuration
$|x_2 - x_1| \sim x_1 \sim x_2 \sim \Lambda ^{-1}$, when  both particles
fly through the nucleus. 

We get the following results:
\bea
\delta A^S &=& \frac {(2\pi)^2}{2\nu^2\;m^2\Lambda ^2} 
{\rm Re}\;\left \{  
\left (\frac {4\Lambda ^2}{m ^2} \right )^{i\nu} 
\frac {\Gamma (1+i\nu)^4}{\Gamma (2+2i\nu)}\;   \Phi _1 (\nu)
\right \},
\\
\delta A^T &=&
\frac {(2\pi)^2}{2 \nu^2\; \Lambda ^2} 
{\rm Re}\;\left \{\frac {(1+i\nu)}{i\nu} 
\left (\frac {4 \Lambda ^2}{m ^2} \right )^{i\nu} 
\frac {\Gamma (1+i\nu)^4}{\Gamma (2+2i\nu)}\; \Phi _1 (\nu)
+\Phi _2 (\nu)
\right \},
\eea
where
\bea
\Phi _1 (\nu) &=& \int \limits _0^{\infty}{\rm d}x x^{1-2i\nu} 
\left \{ 1- e^{-2i\nu K_0(x)} \right \},
\label {phi1}
\\
\Phi _2 (\nu) &=& \frac {1}{(2\pi)^2} 
\int \frac {{\rm d}^2 \vx _1 {\rm d}^2 \vx _2}{(\vx _1-\vx_2)^2} 
\left (\frac {x_1^2}{x_2^2} \right )^{i\nu}
\left \{1-e^{2i\nu \left (K_0(x_1) - K_0(x_2) \right )} \right \}.
\label {phi2}
\eea

Though the function $\Phi _1(\nu)$ can be integrated numerically 
without much problems, the  representation for the function
$\Phi _2(\nu)$, presented above, is not well suited for this purpose.
Moreover, formally, the expression for $\Phi_2(\nu)$, as presented
above, is not well--defined and should be understood within
{\em analytical regularization}.
An integral representation for $\Phi _2 (\nu)$, 
which respects this property and is suitable for numerical
integration, is derived in the Appendix.
The plot of the function $\Phi_2(\nu)$ is presented in Fig.~2.

\begin{figure}
\begin{center}
    \leavevmode
    \epsfxsize=8.cm
    \epsffile[100 250 540 530]{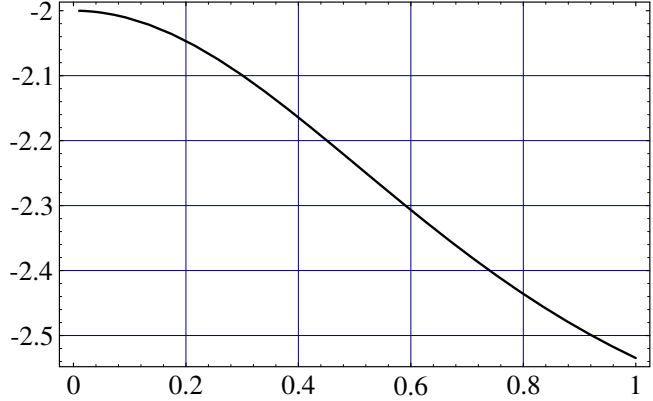}
    \hfill
    \parbox{14.cm}{\small\bf
    \caption[]{\label{figxi4}\sloppy
     {\it The function $\Phi _2 (\nu)$ in dependence on $\nu$.} }}
  \end{center}
\end{figure}

Using Eq. (\ref {sigma2}), we arrive at the 
${\cal O}{(m ^2/\Lambda ^2)}$
correction
to the total cross section of
$\gamma A \to l^+ \l^- A$:
\be
\delta \sigma = Z^2\alpha r_e^2\; \frac {8m^2}{3\Lambda ^2}\;
\frac {1}{\nu^2} {\rm Re} 
\left \{
\left (\frac {2+5i\nu}{2 i\nu} \right ) \left (\frac 
{4\Lambda ^2}{m^2}\right )^{i\nu} \frac {\Gamma (1+i\nu)^2}{\Gamma (2+2i\nu)}
\Phi _1(\nu)+
\Phi _2(\nu) 
\right \}.
\ee

Taking the values $\Lambda = 30\; {\rm MeV}$ and $m = 0.5\; {\rm MeV}$,
we estimate the size of the correction to the case of pure Coulomb potential.
Taking, for instance, $\omega _\gamma = 1\; {\rm Gev}$, we find that the
relative correction does not exceed the $-0.3 $ percent
level for all values of $\nu$.
Moreover, the $\delta \sigma$ decreases more rapidly with the
increase in $\nu$, than the Coulomb cross section does. It is nearly
zero for $\nu \sim 1$.
For this
reason, for larger values of $\nu$, the deviation from Coulomb cross
section is even less important. Hence, we conclude that  
for the electron--positron pair production in the field of a 
heavy nucleus,
one can safely consider the nucleus as a point--like source of 
the Coulomb field. The accuracy of such approximation appears to 
be extremely good.

It should be stressed  however, 
that the reason for the 
smallness of the  correction is mainly related to 
the small ratio $m_e/\Lambda \sim 10^{-2}$.
For the virtual incident photon, the real expansion 
parameter $\mu^2/\Lambda^2 \sim Q^2/\Lambda^2$ can be not so small. 
Simple estimates show that for the virtuality of the 
incident  photon  
$ Q^2 \sim 10~\mbox {MeV}$, the effects related to the 
deviation of the electromagnetic field of the nucleus  
from the Coulomb behavior can become sizeable. 

\subsection{ The case $m \gg \Lambda $}

The opposite limit $\Lambda \ll m$ is interesting for the
muon pair production in the field of a nucleus.
In this case  the resummation is not
necessary because of dipole interaction of the photon
field with the lepton pair. 
Indeed, the form factor $F(k)$ cuts out all photon
momenta $k > \Lambda $. Therefore, the ``remaining''
long--wave photons with $k \le \Lambda$, can not
resolve a dipole formed by a lepton pair, with typical
space separation of the pair components $\sim 1/m$. 
For this reason, any photon
exchange brings in a suppression factor of 
$\Lambda/m$. 

Hence, for
the case $\Lambda \ll m$, it is sufficient
to calculate several lowest order diagrams. Though the resummation is
not necessary in this case, the  representation for the 
scalar and vector impact factors  Eqs. (\ref {scalar},\ref {vector})
appears to be quite useful, in particular for  
the  total cross section.

Without going into any detail of this calculation, we quote the
result for the total cross section in the case 
$\Lambda << m$ including the first ${\cal O}(\Lambda ^2 /m^2)$ correction: 
\be
\sigma = \frac {28}{9} Z^2 \alpha r_e^2 \left \{
\log \left (\frac {2\omega \Lambda}{m^2} \right )
-\frac {171}{42} - \frac {3}{35}\frac {\Lambda ^2}{m^2} 
\left (1 + 8c_1\nu^2 \right ) \right \},~
c_1 = \int \limits_{0}^{\infty} \frac {{\rm d}x}{x^3} 
\left ( 1- xK_1(x) \right )^4 = 0.116.
\label {Lasmall}
\ee

Let us make a remark concerning this result. We would like
to point out, that Eq. (\ref {Lasmall}) provides a reasonable 
approximation to the
scattering cross section
only for very large values of $\omega _\gamma$. 
Indeed, for the values of $\omega_\gamma$,
less than $\omega _{\rm crit} \approx 30m^2/\Lambda$, the cross section
in Eq. (\ref {Lasmall}) is negative. This 
feature is a consequence of the
large negative constant, which accompanies 
the ``large'' logarithm\footnote{We note, that with the logarithmic
accuracy Eq. (\ref {Lasmall}) can be easily 
obtained within equivalent photon approximation.}
and is 
definitely in accord with the dipole screening, which was emphasized
above.

\section {Conclusions}

In this paper, we have considered a lepton pair production
by a high energy photon in the field of the nucleus
using impact representation.
A simple formula for the 
scattering amplitude valid 
to all orders in $Z\alpha$ and arbitrary nuclear potential was derived.  
This representation is close in its form to the known
eikonal representation of scattering amplitudes.

Using this representation,
a simple calculation of the total cross section of the reaction
$\gamma^* A \to l^+ l^- A$ in the limit $\omega _\gamma \gg m, \sqrt {Q^2}$
was demonstrated. 
We hope, that
our results for the pair production by a virtual photon
will be useful for more accurate calculations 
of the pair production in
lepton--nucleus, proton--nucleus or nucleus--nucleus collisions,
than it is usually provided by  Equivalent Photon Approximation.
     
A formula for the amplitude, which is derived in this
paper, is valid irrespective of the exact form of the 
interaction 
potential between the leptons and the  nucleus. For this
reason, our approach allows an easy investigation of the
case of the non--Coulomb potential.  
We have considered
two limits, of strong and weak deviation 
of the interaction potential between lepton
and nucleus from a Coulomb one
and derived the results for the total cross section in these
cases. The case of strong 
deviation is relevant for such processes
as $\gamma A \to \mu ^+ \mu ^- A$ while the case of weak deviation
can be used to describe the $\gamma A \to e^+ e^- A$ process.

Finally, we would like to emphasize the universality of the
obtained results. It is important to realize,
that the formulas for the amplitudes (for 
the scalar and the vector structures),
obtained in this paper, play the same role for high--energy
dipole--charge scattering as the famous eikonal representation
of the scattering amplitudes plays for the charge--charge scattering.
In this respect, the results for the impact factors and methods used in this
paper, definitely have a broader region of applicability, than 
the case of pair production which was considered above. 
In the subsequent paper we are going to demonstrate this by applying
our  technique to the case of the Delbr\"uck scattering.

\section*{Acknowledgments}
We are grateful to R. Kirschner, L. Szymanowski, L.N. Lipatov,
V.S. Fadin and T.T. Wu for useful discussions. 
We are particulary grateful to I.F. Ginzburg who
initiated our collaboration on the problem discussed in this paper.
This research was supported by BMBF under grants  
BMBF-057LP91P0 and  BMBF-057KA92P and 
by Graduiertenkolleg ``Teilchenphysik'' at the University 
of Karlsruhe.
%The research of K.M. was supported by
%BMBF under grant number BMBF-057KA92P
%and by Graduiertenkolleg ``Teilchenphysik'' at the University 
%of Karlsruhe.
%D.I was supported by BMBF-057LP91P0.

\section*{Appendix}

In this Appendix a representation for the function
$\Phi _2 (\nu)$, suitable for numerical integration is derived.
We start from Eq. (\ref {phi2}), where we first integrate over
the relative angle of $\vx _1$ and $\vx _2$.  We then make the change
of variables $x_1 \to x_2 r$, $0 < r < \infty$, $x_2 \to x$
and split the integration region for $r$ into two regions:
 $0 < r < 1$ and $1 < r < \infty$.
Changing
the variable $r \to 1/r$, $x \to xr$ in the second region,
we arrive
at the following result for $\Phi _2 (\nu)$:
\be
\Phi _2 (\nu) = 2 {\rm Re} \left \{ 
\int \limits _{0}^{1} \frac {{\rm d} r\; r^{1+2i\nu}}{1-r^2}
\int \limits _{0}^{\infty} {\rm d}x x
\left (1 - 
e^{2i\nu \left (K_0(xr) - K_0(x) \right )} 
\right )
\right  \}.
\ee

It is further convenient to separate the function 
$\Phi _2 (\nu)$ into two functions in the following manner:
\be
\Phi _2 (\nu) = \Phi _2 ^{(0)}(\nu) + \Phi _2 ^{(1)} (\nu),
\ee
where
\begin{eqnarray}
\Phi _2 ^{(0)}(\nu) &=& 2 {\rm Re} \left \{ 
\int \limits _{0}^{1} {\rm d} r\; r^{1+2i\nu}
\int \limits _{0}^{\infty} {\rm d}x x
\left (1 - 
e^{2i\nu \left (K_0(xr) - K_0(x) \right )} 
\right )
\right  \}.
\\
\Phi _2^{(1)} (\nu) &=& 
 2 {\rm Re} \left \{ 
i\nu \int \limits _{0}^{1} \frac {{\rm d} r\; r^{1+2i\nu}}{1-r^2}
\int \limits _{0}^{\infty} {\rm d}x x
\left [ 1 - e^{2i\nu \left (K_0(x) - K_0(x/r) \right )} \right ] \right  \}.
\end{eqnarray}

The function $\Phi _2^{(1)}$ is already well suited for numerical 
integration. However, this is not the case  for the function
$\Phi _2 ^{(0)}(\nu)$. Furthermore, one notes, that for small values
of $r$, the function $\Phi _2^{(0)}(\nu)$ is proportional
to the integral:
\be
\int \limits _{0}^{r_0}{\rm d}r\; r^{-1+2i\nu}, 
\label{beispiel}
\ee
which is strictly speaking not defined. However, we understand this
integral in a sense of {\em analytical regularization}, which 
allows to set the contribution of the low integration boundary 
in Eq. (\ref  {beispiel}) to zero. 
By the same token, the following identity should be respected
$$
\int \limits_{0}^{\infty} {\rm d}x\; x^{-1+2i\nu} = 0
$$
within the frame of analytical regularization.

With these definitions, the function $\Phi _2^{(0)}(\nu)$ is completely
defined and the following result for it can be obtained:
\be
\Phi _2^{(0)} (\nu) = -2 {\rm Re} \left \{
\int \limits _{0}^{\infty} {\rm d} x x^2 K_1(x) e^{-2i\nu K_0(x)}
\int \limits _{0}^{1} {\rm d} r r^{1+2i\nu} e^{2i\nu K_0(xr)},
\right \}
\ee

For the functions $\Phi_1^{(0)}$ and $\Phi_2^{(0)}$,
written in this form, numerical integration
can be performed directly, without much problems.

%\bibliographystyle{prsty}
%\bibliography{../tex/phd}

\end{document}